\documentclass[conference]{IEEEtran}
\IEEEoverridecommandlockouts
\usepackage{cite}
\usepackage{amsmath,amssymb,amsfonts}
\usepackage{algorithm}
\usepackage{algcompatible}% http://ctan.org/pkg/algorithmicx
\usepackage{float} % 导入float包
\usepackage{graphicx}
\usepackage{textcomp}
\usepackage{xcolor}
\pdfoutput=1

\makeatletter
\newenvironment{breakablealgorithm}
  {% \begin{breakablealgorithm}
   \begin{center}
     \refstepcounter{algorithm}% New algorithm
     \hrule height.8pt depth0pt \kern2pt% \@fs@pre for \@fs@ruled
     \renewcommand{\caption}[2][\relax]{% Make a new \caption
       {\raggedright\textbf{\ALG@name~\thealgorithm} ##2\par}%
       \ifx\relax##1\relax % #1 is \relax
         \addcontentsline{loa}{algorithm}{\protect\numberline{\thealgorithm}##2}%
       \else % #1 is not \relax
         \addcontentsline{loa}{algorithm}{\protect\numberline{\thealgorithm}##1}%
       \fi
       \kern2pt\hrule\kern2pt
     }
  }{% \end{breakablealgorithm}
     \kern2pt\hrule\relax% \@fs@post for \@fs@ruled
   \end{center}
  }
\makeatother

\def\BibTeX{{\rm B\kern-.05em{\sc i\kern-.025em b}\kern-.08em
    T\kern-.1667em\lower.7ex\hbox{E}\kern-.125emX}}
\begin{document}

\title{Dog Heart Rate and Blood Oxygen Metaverse Interaction System}

\author{
\IEEEauthorblockN{1\textsuperscript{st} Yanhui Jiang}
\IEEEauthorblockA{
   \textit{Department of Computer Science} \\
   \textit{University College London (UCL)}\\
   London, WC1E 6BT, UK\\
   yanhui.jiang.23@ucl.ac.uk}
\and
\IEEEauthorblockN{2\textsuperscript{nd*} Jin Cao}
\IEEEauthorblockA{
   \textit{Independent Researcher} \\
   \textit{Johns Hopkins University}\\
   Baltimore, MD, 21218, USA \\
   caojinscholar@gmail.com}
\and
\IEEEauthorblockN{2\textsuperscript{nd} Chang Yu}
\IEEEauthorblockA{
   \textit{Independent Researcher} \\
   \textit{Northeastern University}\\
   Boston, MA, 02115, USA \\
   chang.yu@northeastern.edu}   
}

\maketitle

\begin{abstract}
This study developed an improved dog heart rate and blood oxygen sensor system using Arduino. Traditional methods face accuracy and reliability issues. Our system integrates advanced computational techniques with hardware-based sensing to enhance measurement precision. An Arduino microcontroller connected to a heart rate and blood oxygen sensor collects raw data, which is preprocessed and filtered to remove noise. Experimental plans include long-term monitoring of multiple dogs and comparative analysis with traditional methods to validate the system's effectiveness, aiming for widespread use in the home or veterinary clinics. Additionally, the system offers dog owners the possibility to interact with their virtual dogs in the metaverse using AR/VR, allowing them to better understand their real dogs' health conditions by observing their virtual health status.
\end{abstract}

\begin{IEEEkeywords}
Dog monitoring, Heart rate sensor, Blood oxygen sensor, Data preprocessing, Signal filtering, Metaverse, Computational techniques, Sensor accuracy
\end{IEEEkeywords}

\section{Introduction}

The trend of pet ownership among Generation Z is reshaping social dynamics and relationships\cite{b1}. According to recent statistics, 76\% of Generation Z prefers pets over children, spending more than \$300 billion on pet-related expenses\cite{b4}. This significant shift is highlighted by the increase in pet ownership from 56\% in 1988 to 70\% by 2023\cite{b7}. This phenomenon suggests that pets are increasingly viewed as alternatives to family members, fulfilling emotional and social needs traditionally met through human interaction\cite{b6}.

This shift is driven by both intrinsic and extrinsic factors. Intrinsically, Generation Z is highly goal-oriented and focused on personal growth\cite{b9}. They invest considerable energy and resources into pet care, viewing pets as essential to their lives and personal development\cite{b3}. Extrinsically, pets provide significant psychological benefits, such as reduced social anxiety and improved social skills\cite{b10}. The presence of pets encourages everyday behaviors and responsibility and positively contributes to interpersonal communication and overall mental health. Despite these benefits, this trend poses potential challenges to individual and societal well-being\cite{b5}. Observations suggest that Generation Z increasingly favors pets over traditional family relationships, which may have adverse effects on mental health and social interactions, potentially undermining social cohesion\cite{b2, b8}. Additionally, Owners can use AR/VR to interact with virtual dogs in the metaverse, understanding real dogs' health by observing virtual health status.

We made an initial Draft of the Design and Scenario Definition of the Dog Heart Rate Blood Oxygen Sensor Detection System below:

\begin{figure}[H]  % 更改环境名称为figure，并且添加位置参数
    \centering
    \includegraphics[width=\columnwidth]{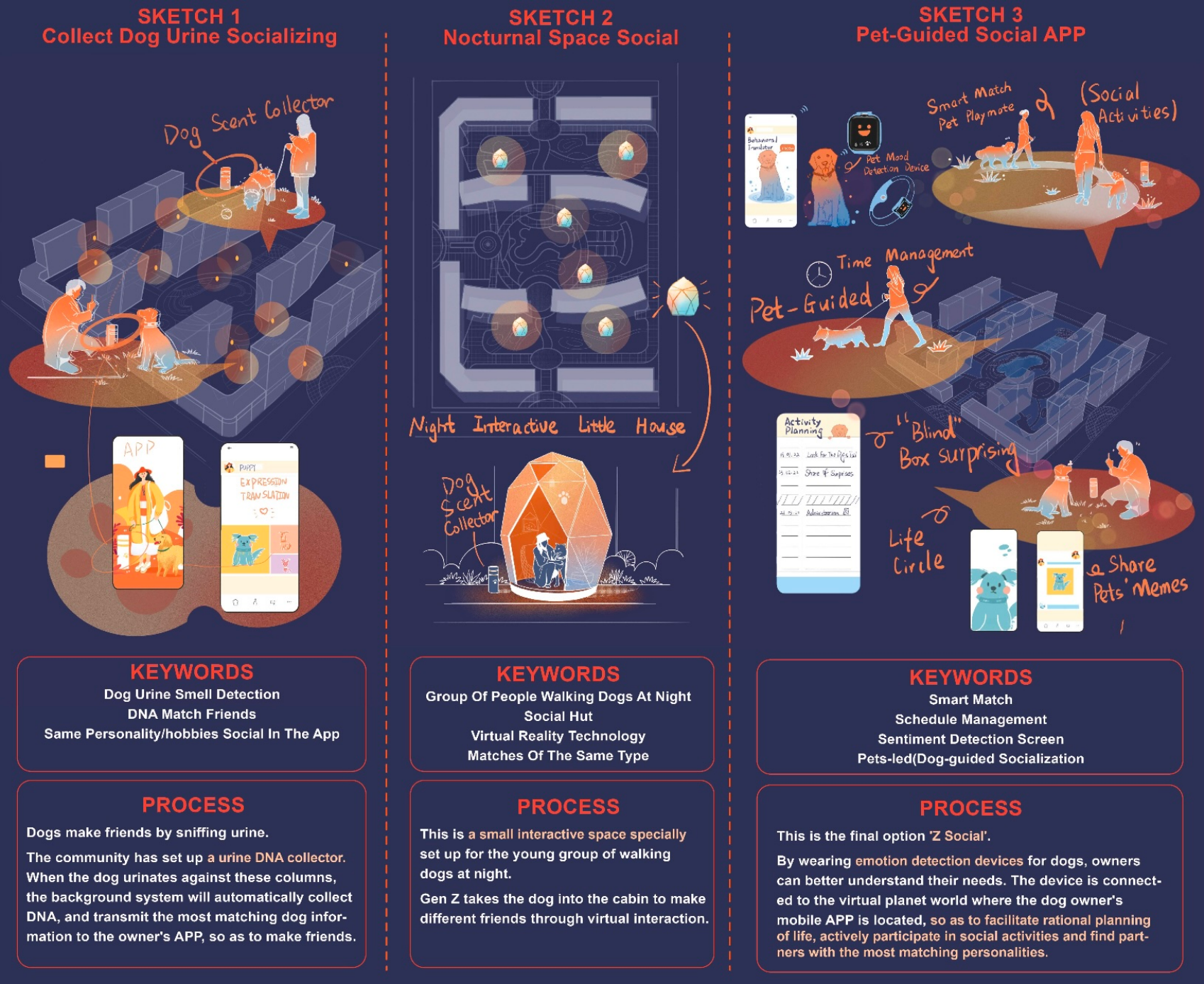}  % 改为列宽
    \caption{Initial Draft on the Design and Scenario Definition of the Dog Heart Rate Blood Oxygen Sensor Detection System.}
    \label{fig1}
\end{figure}

Additionally, we have introduced an improved dog heart rate and blood oxygen sensor detection system using Arduino system in future. An Arduino microcontroller is connected to heart rate and blood oxygen sensors, and the raw data collected by the sensors undergoes preprocessing and filtering to eliminate noise and outliers. Subsequently, rough set theory \cite{b47} is applied to further refine and analyze the information in future research, identifying significant patterns and correlations. This approach not only improves data accuracy but also enhances the system's robustness, providing an effective tool for pet health monitoring.

The rest of this paper is organized as follows. Section II discusses related work, emphasizing recent years' new tech that influenced my topic, such as computational advances in veterinary medicine, precision in data analysis, and wearable technology's impact on animal healthcare. Section III presents the equations and algorithms used to enhance our system's data accuracy and processing capabilities. Section IV explores the interaction journey between dogs and humans, informing the design of our system. Section V details the prototype design, focusing on user experience and material selection. Section VI outlines our evaluation plan, describing the system testing, metaverse application and data analysis phases. Section VII discusses future directions, including the integration of additional sensors and enhanced mobile app features. Section VIII concludes by summarizing our findings and the potential impact of our system on pet health monitoring.

\section{Related Work}
We summarized related work on our topic of dog heart rate blood oxygen sensor detection system. We also extended the following 3 topics to better understand the current technology that influences our research:

\subsection{Computational Advances in Veterinary Medicine}
The significance of animals in contemporary society extends beyond companionship, encompassing crucial roles in ecological balance and human well-being. Drawing parallels from human medical practices, veterinary medicine has also undergone a transformation, increasingly incorporating computational technologies to enhance diagnostic and therapeutic capabilities. The integration of sophisticated computing techniques in animal healthcare mirrors advancements in human medicine, particularly in respiratory rate\cite{b44}, face image computing\cite{b51}, and health monitoring systems. In my topic, better understanding a dog's emotions or status is one of the goals for this system, breathing detection\cite{b43}, robot control\cite{b57}, mental stress\cite{b42}, and temperature measurements\cite{b41} are important to help improve the system. These advancements have paved the way for more precise and less invasive methods to assess health conditions, illustrating the convergence of veterinary and human medical sciences through shared technological innovations.

\subsection{Precision in Data Analysis Through Computational Models}
In the realm of both human and veterinary medicine, the accuracy of data interpretation is paramount, especially when dealing with ambiguous or complex datasets. In recent years, many new technologies have influenced this area. The computing methods can enhance the accuracy of image data, such as federated learning\cite{b33, b34}, multi-focus\cite{b21} and multi-modal\cite{b36} image recognition, which can improve human or pets image recognition. Communication accuracy between humans and pets is also important, and the related recent technology of enhanced large language/text models\cite{b17, b25} helps this process. The development of precise computational models has become a cornerstone in various industries, facilitating enhanced data analysis capabilities. These models leverage advanced algorithms and machine learning techniques to refine data accuracy, thereby supporting critical decision-making processes in clinical settings. In veterinary applications, such as the monitoring of heart rate and blood oxygen levels in dogs, these models play a vital role in ensuring the reliability of the diagnostics, thus directly impacting treatment outcomes and animal welfare.

\subsection{Wearable Devices and Identification Systems}
Wearable technology has revolutionized the way health monitoring and diagnostics are approached in medical sciences. Identification and localization features, commonly utilized in other sectors such as security and mobile communications, have been adapted to enhance the functionality of these wearable devices in animal healthcare. For the identification system after pets wear it, the essence lies in motion detection. In recent years, trajectory motion computing technologies have flourished, such as visual SLAM\cite{b28}, visual prediction\cite{b29} and enhancement\cite{b56}.This integration not only facilitates continuous health monitoring but also aids in the development of systems capable of early disease detection and management, thereby contributing to broader research applications and improved healthcare solutions for animals.

\subsection{Other ML/AI computing research can influence my intelligent system design}
Other new improvement research in engineering artificial intelligence and machine learning in peripheral industries have also contributed to advancements. These advancements provide valuable references and insights for optimizing data and achieving precise recognition results in my project. For example, among the other breakthroughs in engineering artificial intelligence and machine learning, such as HDRTV network improvement\cite{b18}, enhanced\cite{b24} Reinforcement\cite{b30} and other types of machine learning\cite{b55}, the artificial intelligence in financial\cite{b13, b32}, security prediction\cite{b11} and recommendation system\cite{b15}, and biomedical image processing\cite{b58}.

These sections aim to connect the technological advancements in computing and their application in veterinary medicine, highlighting the synergy between human and animal healthcare technologies and the critical role of precision in data-driven diagnostics. These related works have a good influence to our research, and let us deeper jumping into our topic with good detection system building.

\section{EQUATIONS AND ALGORITHM}

This paper proposes an improved dog heart rate and blood oxygen sensor detection system that combines Arduino hardware. The system collects data on the dog's heart rate and blood oxygen saturation using sensors, and analysis to enhance accuracy and reliability. The specific algorithm is as follows:

Data Collection: Real-time collection of the dog's infrared (IR) and red light (Red) reflection values using the MAX30102 sensor controlled by Arduino. These values fluctuate with changes in heartbeats and blood oxygen levels.

Heart Rate Detection: Detect heartbeats based on the collected IR values by identifying significant pulse signals (notable increases or decreases in IR values). Record the timestamp of each heartbeat and calculate the interval (\(\Delta t\)) between consecutive heartbeats. Use the formula \(\text{BPM} = \frac{60}{\Delta t}\) to calculate beats per minute (BPM) and take the average over multiple beats to enhance measurement stability.

Blood Oxygen Detection: Calculate the ratio (\(\text{Ratio}\)) of the collected red light values to the IR values. Use the empirical formula \(\text{SpO2} = A - B \times \text{Ratio}\) to calculate blood oxygen saturation (SpO2) and ensure its value is within a reasonable range (0\% to 100\%).

Results Output: The system displays the dog's heart rate and blood oxygen saturation in real time and provides an assessment of emotional states. By analyzing heart rate and body temperature data combined with rough set theory, the system determines the dog's emotional state.

\subsection{Equations 1: MAX30102 sensor Heart Rate and Blood Oxygen Detection}

Input: Raw Heart Rate and Blood Oxygen sample data

Output: Processed Heart Rate and Blood Oxygen data

\subsubsection{Heart Rate Detection Algorithm}

For IR Value Acquisition, here we use the sensor reads infrared (IR) light reflection values. Each heartbeat causes changes in blood volume, altering the IR light reflection.

(1a). Heartbeat Detection: Detect heartbeats by identifying significant changes in IR values. Given a series of IR values \(\{IR_1, IR_2, \ldots, IR_n\}\), a heartbeat is detected by significant changes in IR values. Let's denote the times when heartbeats are detected as \( t_i \).

(1b). Heartbeat Interval Calculation: Compute the time interval \(\Delta t\) between consecutive heartbeats:
   \[
   \Delta t = t_{i} - t_{i-1}
   \]

(1c). Heart Rate Calculation: Convert the time interval to beats per minute (BPM).The number of heartbeats that occur in one minute (60 seconds). By dividing 60 by the time taken for a single heartbeat, you can obtain the beats per minute (BPM), where \(\Delta t\) is in seconds:
   \[
   \text{BPM} = \frac{60}{\Delta t}
   \]

(1d). Average Heart Rate Calculation: To stabilize the heart rate measurement, calculate the average heart rate over a set of recent beats. Given an array of recent BPM measurements \(\{BPM_1, BPM_2, \ldots, BPM_N\}\), the average heart rate \(\overline{BPM}\) is:
   \[
   \overline{BPM} = \frac{1}{N} \sum_{i=1}^{N} BPM_i
   \]

\subsubsection{Blood Oxygen Detection Algorithm}

(2a). Red and IR Value Acquisition: The sensor reads both red and infrared (IR) light reflection values.

(2b). Ratio Calculation: Calculate the ratio of red light value to IR light value. Given red light values \( R \) and IR light values \( IR \):
   \[
   \text{Ratio} = \frac{R}{IR}
   \]

(2c). Blood Oxygen Saturation Calculation: Use the ratio to estimate the blood oxygen saturation level (SpO2), with \(A\) and \(B\) being constants derived from experimental data and subject to sensor calibration. This formula is derived from experimental data and may vary based on sensor calibration:
  \[
   \text{SpO2} = A - B \times \text{Ratio}
   \]
   
(2d). SpO2 Range Limitation: Ensure the calculated SpO2 is within a reasonable range (0\% to 100\%), while $ \min(100, \text{SpO2}) $  ensures that the SpO2 value does not exceed 100\%. If the calculated SpO2 is greater than 100\%, it is capped at 100\%, and $\max(0, \min(100, \text{SpO2}))$  ensures that the SpO2 value is not less than 0\%. After applying the previous step, if the SpO2 is still less than 0\%, it is set to 0:
   \[
   \text{SpO2} = \max(0, \min(100, \text{SpO2}))
   \]

\subsection{Algorithm 1: MAX30102 sensor Heart Rate and Blood Oxygen Detection}

Algorithm 1: This algorithm, implemented using Arduino and sensors, detects heart rate and blood oxygen saturation. Initially, the sensor is initialized and configured, then an infinite loop begins. Within the loop, the sensor reads infrared (IR) and red light values. For heart rate detection, the algorithm identifies heartbeats by detecting significant changes in IR values, calculates the time intervals between consecutive heartbeats, and derives beats per minute (BPM). It updates the average heart rate. If the IR value is above a threshold, the current and average BPM are displayed. 

Here the input data is Raw Heart Rate and Blood Oxygen sample data, and the Output data is Processed Heart Rate and Blood Oxygen data.

\begin{breakablealgorithm}
\caption{Heart Rate and Blood Oxygen Detection}
\label{alg:heart_rate_oxygen}
\begin{algorithmic}[1]
\STATE Initialize sensor
\STATE Configure sensor settings

\WHILE{true}
    \STATE // Heart Rate Detection
    \STATE Read IR value from sensor
    \IF{heartbeat detected}
        \STATE currentTime = get current time
        \STATE deltaTime = currentTime - lastBeatTime
        \STATE lastBeatTime = currentTime
        
        \STATE BPM = 60 / (deltaTime / 1000.0)
        
        \IF{BPM in valid range}
            \STATE Store BPM in heart rate array
            \STATE Update average BPM from heart rate array
        \ENDIF
    \ENDIF
    
    \IF{IR value above threshold}
        \STATE Display current BPM and average BPM
    \ELSE
        \STATE Display ``No finger detected''
    \ENDIF
    
    \STATE // Blood Oxygen Detection
    \STATE Read red value from sensor
    \STATE Read IR value from sensor
    \STATE ratio = red value / IR value
    \STATE SpO2 = A - B * ratio
    \STATE SpO2 = constrain(SpO2, 0, 100)  // Ensure SpO2 is within 0\% to 100\%
    \STATE Display SpO2
    
    \STATE WAIT for 1000 milliseconds
\ENDWHILE

\end{algorithmic}
\end{breakablealgorithm}

\section{Dog and human interaction journey research}
Before developing the Dog Heart Rate Blood Oxygen Sensor Detection System, our team conducted an in-depth study of the interaction journey between dogs and their owners. By conducting a series of interviews with dog owners, we collected detailed information about their daily interactions with their pets to establish a comprehensive interaction journey map. Here is the conclusion chart of the dog and human user journey map below:

\begin{figure}[H]
    \centering
    \includegraphics[width=\columnwidth]{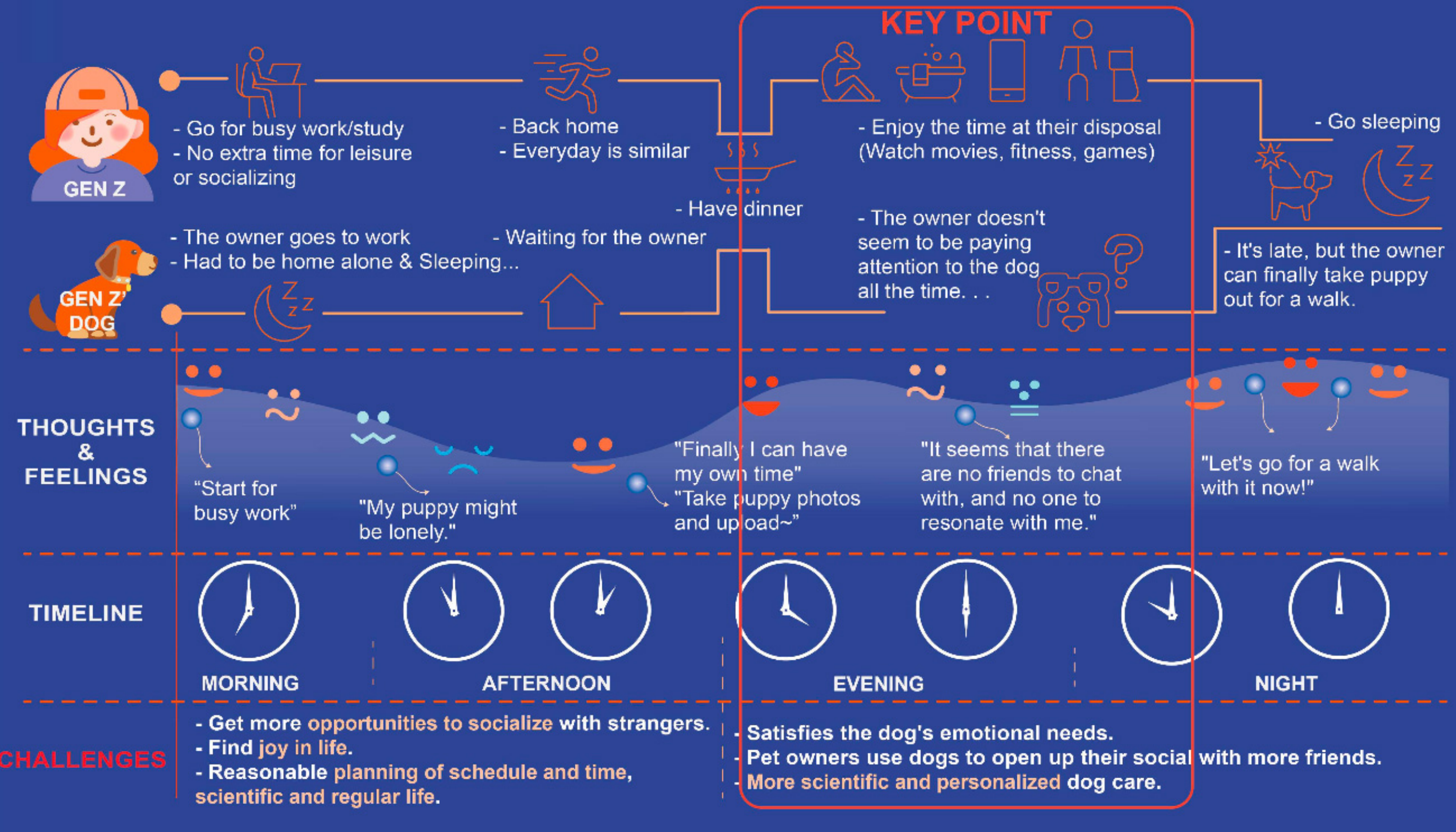}
    \caption{The dog and human user journey map.}
    \label{fig3}
\end{figure}

Through detailed research into the dog-human interaction journey, we laid a solid foundation for the development of the dog heart rate blood oxygen sensor detection system. This system not only elevates the technological level of pet health monitoring but also provides strong support for enhancing pet welfare and promoting positive interactions between pets and their owners. 

Additionally, we plan to enable the system to offer pet owners the possibility to interact with virtual dogs in the metaverse using AR/VR in the future. By observing the virtual dogs' health status, pet owners can better understand their real dogs' health conditions. This feature not only enhances pet owners' understanding of their pets but also allows them to experience more interactive fun with their pets in a virtual environment, thereby managing their pets' health and emotional needs more comprehensively.

\section{Prototype design and building}

Based on dog and human research, we understand the outlook, fitting, and material needs from dogs, and make the following prototype design and building.

Enclosure Design: A dog collar's enclosure design is critical in balancing functionality and comfort. As shown in the sketch, the iterative design process highlights a process of optimizing a collar that is the least intrusive to the dog's natural behavior. The final design solution features a streamlined collar focusing on ergonomics and ease of use. This design ensured that the LCD showing the dog's emotional state was in a position that was easy for the owner to read without causing discomfort to the dog. 

Colour scheme: The collar features a neutral, modern color scheme with black, white, and blue primary tones. This color palette was chosen for aesthetic reasons and to ensure that the collar is unobtrusive and sophisticated. The use of soft black silicone and leather elements enhances the stylish look of the
collar whilst also improving overall comfort. The combination of these hues not only adds a touch of elegance but also avoids overwhelming the natural beauty of the dog, integrating smoothly with various settings, whether at home or out in public. Here is the experimented sketch and modeling for the enclosure design and structure:

\begin{figure}[H]
    \centering
    \includegraphics[width=\columnwidth]{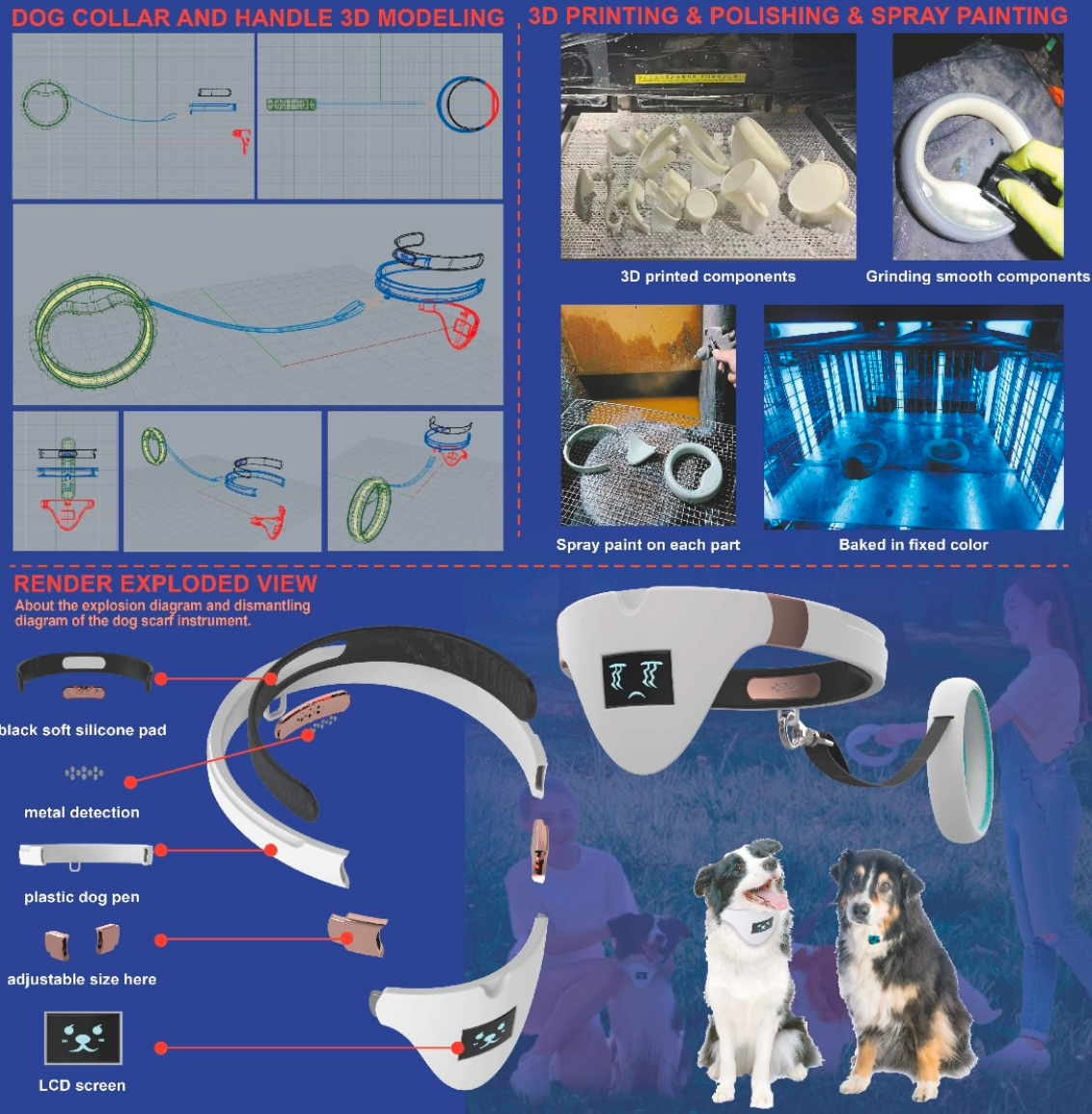}
    \caption{The experimented sketch and modeling for the enclosure design and structure}
    \label{fig4}
\end{figure}

\section{EVALUATION PLAN}

The combination of advanced computational techniques and robust hardware design ensures that the collar is reliable for monitoring a dog's health. 

The collar has built-in sensors to monitor your dog's heart rate and body temperature. The heart rate sensor is strategically placed near the chest to accurately detect changes in blood volume. 

When integrating sensor hardware within the collar, meticulous attention to detail is paid to ensure functionality and comfort. 

Here is the sensor listed we are using and connected in the evaluation plan:

\begin{figure}[H]
    \centering
    \includegraphics[width=\columnwidth]{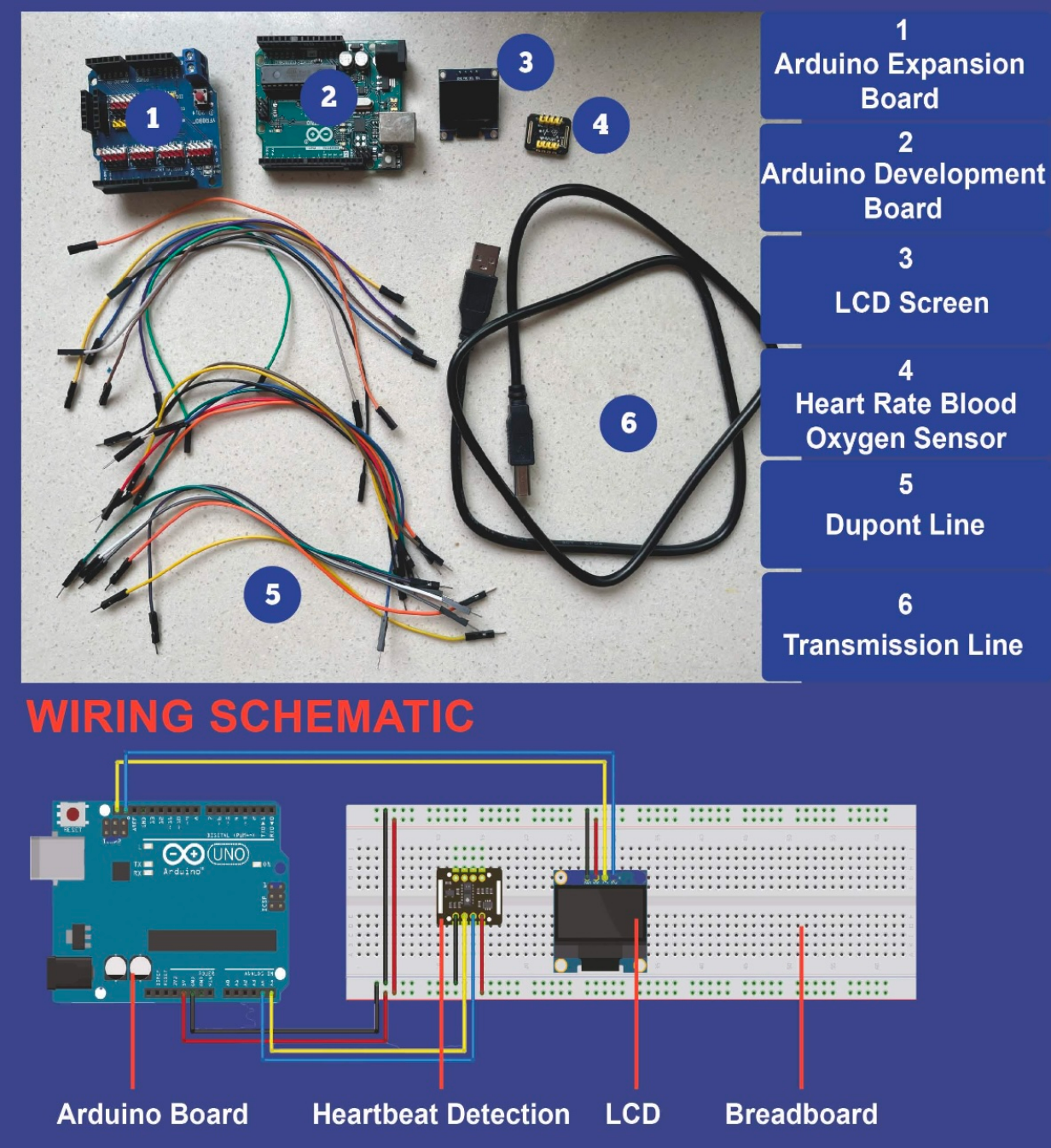}
    \caption{Sensor listed are using and connected in the evaluation plan}
    \label{fig5}
\end{figure}

\subsection{Experimental process and plan.}

The depicted experimental process involves a collaborative effort by our team to develop and test a wearable dog collar that integrates sensors for monitoring physiological signals. Initially, our team sets up and programs the Arduino-based sensors, ensuring they are finely calibrated to accurately capture the data. 

Here is the plan for the experiment process, you can see the example of our team's Arduino testing experiment process and evaluation details below:

\begin{figure}[H]
    \centering
    \includegraphics[width=\columnwidth]{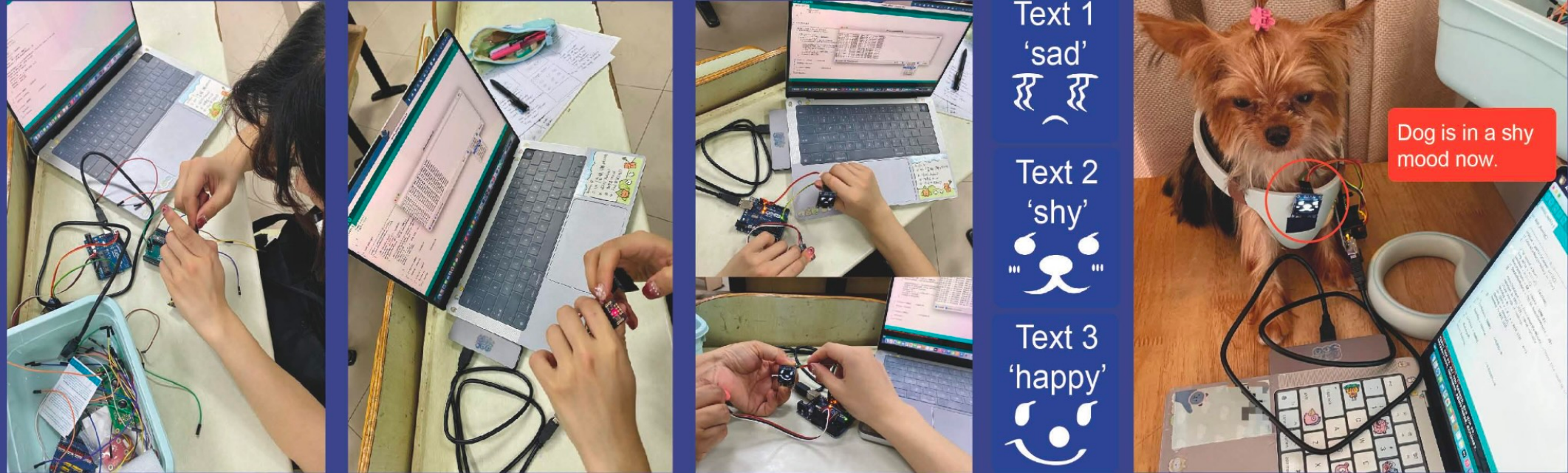}
    \caption{The plan of Arduino sensors' experiment process}
    \label{fig6}
\end{figure}

Through the user interaction satisfaction testing, an example of this is the GSM-HEART test, which outlines the user goal, user interaction signals, metrics, and the HEART model—Happiness, Engagement, Adoption, Retention, Task success. This model helps to ascertain whether the designed system fulfills the user's needs. The findings from this model provide insights into areas where the application may need further refinement to enhance user satisfaction and overall effectiveness. An example of this is the GSM-HEART below:

\begin{figure}[H]
    \centering
    \includegraphics[width=\columnwidth]{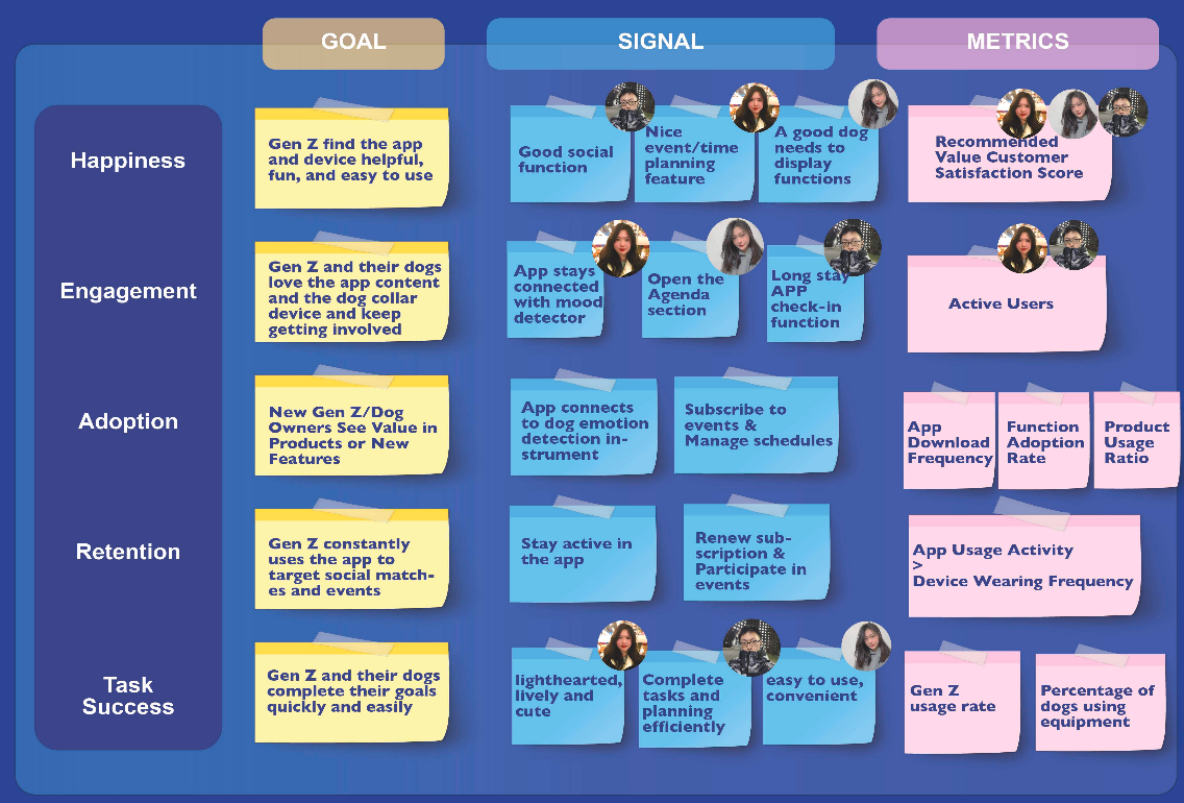}
    \caption{The GSM-Heart testing example}
    \label{fig7}
\end{figure}

\subsection{Metaverse application plan}

In the future, we plan to enable the system to offer pet owners the possibility to interact with virtual dogs in the metaverse using AR/VR. By observing the virtual dogs' health status, pet owners can better understand their real dogs' health conditions. Furthermore, the integration of AR/VR technology can provide a new dimension to pet care, making it more engaging and insightful. This virtual interaction platform can also serve as a valuable tool for educating pet owners on best practices for pet health and well-being, ultimately fostering a deeper bond between pets and their owners. The metaverse component will include features such as virtual health dashboards, interactive training modules, and social spaces for pet owners to connect and share experiences.

\section{FUTURE DISCUSSION}

As we move forward with the development of the Dog Heart Rate Blood Oxygen Sensor Detection System, our team is dedicated to expanding and refining the capabilities of this innovative tool. Here are the key areas we aim to address in our future plans:

Integration of Additional Sensors: To provide a more comprehensive health monitoring solution, we plan to integrate additional sensors into the dog collar. These will include sensors for measuring things like stress levels, calorie expenditure, and more precise activity tracking.

Expansion to Other Platforms: To reach a wider audience, we will develop versions of the app suitable for multiple platforms, including iOS, Android, web-based interfaces, and all types of metaverse platforms and applications. This will ensure that all pet owners have access to our tools, regardless of their device preferences.

\section{CONCLUSION}
This research explores the prototype of the Dog Heart Rate Blood Oxygen Sensor Detection System, aimed at improving dog-owner interaction through advanced technology. We developed a plan to test integrated sensor technology for monitoring heart rate and blood oxygen levels. The system needs further testing and optimization, focusing on sensor integration, data processing precision, and monitoring capabilities. Future plans include adding sensors for stress and activity levels to create a comprehensive pet health monitoring system. Additionally, the system plans to integrate metaverse technology, allowing pet owners to interact with virtual dogs through AR/VR, thereby gaining a deeper understanding and engagement in their pets' health management.


\begin{thebibliography}{00}

\bibitem{b1} Katharine L. Anderson, Rachel A. Casey, Ben Cooper, Melissa M. Upjohn, and Robert M. Christley, "National Dog Survey: Describing UK Dog and Ownership Demographics." \textit{Animals: an Open Access Journal from MDPI}, vol. 13, no. 6, pp. 1072, 2023.

\bibitem{b2} Thomas Astell-Burt, Terry Hartig, I Gusti Ngurah Edi Putra, Ramya Walsan, Tashi Dendup, and Xiaoqi Feng, ``Green space and loneliness: A systematic review with theoretical and methodological guidance for future research,'' \textit{Science of The Total Environment}, vol. 847, pp. 157521, 2022.

\bibitem{b3} Bhagyashree Barhate and Khalil M. Dirani, ``Career aspirations of generation Z: a systematic literature review,'' \textit{European Journal of Training and Development}, vol. 46, no. 1/2, pp. 139--157, 2021.

\bibitem{b4} Taryn Graham, Katrina Milaney, Cindy Adams, and Melanie Rock, ``Are Millennials really Picking Pets over People? Taking a Closer Look at Dog Ownership in Emerging Adulthood,'' \textit{Canadian Journal of Family and Youth / Le Journal Canadien de Famille et de la Jeunesse}, vol. 11, pp. 202--227, 2019.

\bibitem{b5} Catarina Martins, Jorge Soares, António Cortinhas, Luís Silva, Luís Cardoso, Maria Pires, and Maria Mota, ``Pet’s influence on humans’ daily physical activity and mental health: a meta-analysis,'' \textit{Frontiers in Public Health}, vol. 11, pp. 1196199, 2023.

\bibitem{b6} Allen R. Mcconnell, Christina M. Brown, Tonya M. Shoda, L. Stayton, and Colleen E. Martin, ``Friends With Benefits: On the Positive Consequences of Pet Ownership,'' \textit{Journal of Personality and Social Psychology}, 2011. Retrieved May 19, 2024.

\bibitem{b7} Renata Roma, Christine Tardif-Williams, Shannon Moore, and Patricia Pendry, ``My ‘Perfect’ Dog: Undesired Dog Behaviours and Owners’ Coping Styles,'' \textit{Human-Animal Interactions}, hai.2023.0011, 2023.

\bibitem{b8} ``A person-organization fit Model of Generation Z: Preliminary studies,'' \textit{Journal of Entrepreneurship, Management and Innovation}, vol. 16, no. 4, pp. 149--176, 2020.

\bibitem{b9} ``What to know about Gen Z,'' Retrieved May 19, 2024 from \textit{Stanford News}, 2022.

\bibitem{b10} "The power of support from companion animals for people living with mental health problems: a systematic review and narrative synthesis of the evidence," \textit{BMC Psychiatry}, 2018. Retrieved May 19, 2024 from \textit{Springer}.

\bibitem{b11} Mo, Yuhong, Shaojie Li, Yushan Dong, Ziyi Zhu, and Zhenglin Li. "Password Complexity Prediction Based on RoBERTa Algorithm." \textit{Applied Science and Engineering Journal for Advanced Research} 3, no. 3 (2024): 1-5.

\bibitem{b13} Li, Zhenglin, Hanyi Yu, Jinxin Xu, Jihang Liu, and Yuhong Mo. "Stock market analysis and prediction using LSTM: A case study on technology stocks." \textit{Innovations in Applied Engineering and Technology} (2023): 1-6.

\bibitem{b15} Liu, Tianrui, Changxin Xu, Yuxin Qiao, Chufeng Jiang, and Weisheng Chen. "News recommendation with attention mechanism." \textit{arXiv preprint arXiv:2402.07422} (2024).

\bibitem{b17} He, Gang, Kepeng Xu, Li Xu, Chang Wu, Ming Sun, Xing Wen, and Yu-Wing Tai. "SDRTV-to-HDRTV via hierarchical dynamic context feature mapping." In \textit{Proceedings of the 30th ACM International Conference on Multimedia}, pp. 2890-2898, 2022.

\bibitem{b18} Xu, Kepeng, Gang He, Li Xu, Xingchao Yang, Ming Sun, Yuzhi Wang, Zijia Ma, Haoqiang Fan, and Xing Wen. "Towards Robust SDRTV-to-HDRTV via Dual Inverse Degradation Network." \textit{arXiv preprint arXiv:2307.03394} (2023).

\bibitem{b21} Gai, Di, Xuanjing Shen, Haipeng Chen, and Pengxiang Su. "Multi-focus image fusion method based on two stage of convolutional neural network." \textit{Signal Processing} 176 (2020): 107681.

\bibitem{b24} Song, J., Liu, H., Li, K., Tian, J., \& Mo, Y. (2024). "A Comprehensive Evaluation and Comparison of Enhanced Learning Methods." \textit{Academic Journal of Science and Technology} 10(3), 167-171.

\bibitem{b25} Liu, T., Li, S., Dong, Y., Mo, Y., \& He, S. (2024). "Spam Detection and Classification Based on DistilBERT Deep Learning Algorithm." \textit{Applied Science and Engineering Journal for Advanced Research} 3(3), 6-10.

\bibitem{b28} Jin, S., Dai, X., Meng, Q. "‘Focusing on the right regions’—Guided saliency prediction for visual SLAM." \textit{Expert Systems with Applications}, 213 (2023): 119068.

\bibitem{b29} Jin, S., Dai, X., Meng, Q. "Loop closure detection with patch-level local features and visual saliency prediction." \textit{Engineering Applications of Artificial Intelligence}, 120 (2023): 105902.


\bibitem{b30} Chen, J., Mao, C., Sha, G., Sheng, W., Fan, H., Wang, D., ... \& Zhang, Y. (2024). "Reinforcement learning based two‐timescale energy management for energy hub." \textit{IET Renewable Power Generation} 18(3), 476-488.

\bibitem{b32} Chen, J., Mao, C., Wang, D., Qiu, S., Ma, C., \& Liu, Z. (2023, March). "Robust Optimization Based Multi-level Coordinated Scheduling Strategy for Energy Hub in Spot Market." In \textit{2023 7th International Conference on Green Energy and Applications (ICGEA)}, pp. 59-65. IEEE.

\bibitem{b33} Huang, Wenke, Guancheng Wan, Mang Ye, and Bo Du. "Federated graph semantic and structural learning." In \textit{Proc. Int. Joint Conf. Artif. Intell}, pp. 3830-3838, 2023.

\bibitem{b34} Wan, Guancheng, Wenke Huang, and Mang Ye. "Federated Graph Learning under Domain Shift with Generalizable Prototypes." In \textit{Proceedings of the AAAI Conference on Artificial Intelligence}, vol. 38, no. 14, pp. 15429-15437, 2024.

\bibitem{b36} Wang, J., Li, X., Jin, Y., Zhong, Y., Zhang, K., \& Zhou, C. (2024). "Research on Image Recognition Technology Based on Multimodal Deep Learning." \textit{arXiv preprint arXiv:2405.03091}.


\bibitem{b41} Yao, C., Chen, H., Onishi, T., Datta-Gupta, A., Mawalkar, S., Mishra, S., \& Pasumarti, A. (2021, September). "Robust CO2 Plume Imaging Using Joint Tomographic Inversion of Distributed Pressure and Temperature Measurements." In \textit{SPE Annual Technical Conference and Exhibition}, p. D021S023R006. SPE.

\bibitem{b42} Cho, Youngjun, Simon J. Julier, and Nadia Bianchi-Berthouze. "Instant stress: detection of perceived mental stress through smartphone photoplethysmography and thermal imaging." \textit{JMIR mental health} 6.4 (2019): e10140.

\bibitem{b43} Cho, Youngjun, Nadia Bianchi-Berthouze, and Simon J. Julier. "DeepBreath: Deep learning of breathing patterns for automatic stress recognition using low-cost thermal imaging in unconstrained settings." In \textit{2017 Seventh international conference on affective computing and intelligent interaction (acii)}, IEEE, 2017.

\bibitem{b44} Cho, Y., Julier, S. J., Marquardt, N., \& Bianchi-Berthouze, N. (2017). "Robust tracking of respiratory rate in high-dynamic range scenes using mobile thermal imaging." \textit{Biomedical optics express} 8(10), 4480-4503.

\bibitem{b47} Maji, Pradipta. "A Rough Hypercuboid Approach for Feature Selection in Approximation Spaces." \textit{IEEE Transactions on Knowledge and Data Engineering} 26, no. 1 (2012): 16-29. doi: 10.1109/TKDE.2012.231.

\bibitem{b51} Y. Lai, G. Yang, Y. He, Z. Luo, and S. Li, ``Selective Domain-Invariant Feature for Generalizable Deepfake Detection,'' in \textit{Proc. IEEE International Conference on Acoustics, Speech and Signal Processing (ICASSP)}, April 2024, pp. 2335--2339.

\bibitem{b55} Chen, Z., Ge, J., Zhan, H., Huang, S., \& Wang, D. (2021). Pareto self-supervised training for few-shot learning. In Proceedings of the IEEE/CVF conference on computer vision and pattern recognition (pp. 13663-13672).

\bibitem{b56} Liu, J., Bu, Y., Tso, D., \& Qiu, Q. (2023, October). Improved Efficiency Based on Learned Saccade and Continuous Scene Reconstruction From Foveated Visual Sampling. In The Twelfth International Conference on Learning Representations.

\bibitem{b57} Z. Lin and F. Xu, "Simulation of Robot Automatic Control Model Based on Artificial Intelligence Algorithm," 2023 2nd International Conference on Artificial Intelligence and Autonomous Robot Systems (AIARS), Bristol, United Kingdom, 2023, pp. 535-539, doi: 10.1109/AIARS59518.2023.00113.

\bibitem{b58} Jiang, C., Hou, X., Kondepudi, A., Chowdury, A., Freudiger, C. W., Orringer, D. A., ... \& Hollon, T. C. (2023). Hierarchical discriminative learning improves visual representations of biomedical microscopy. In Proceedings of the IEEE/CVF Conference on Computer Vision and Pattern Recognition (pp. 19798-19808).


\end{thebibliography}
\end{document}